\documentclass[10pt,a4paper,journal,twocolumn]{IEEEtran}
\usepackage{amsmath,cite,graphicx,amssymb}
\title{Fundamental Imaging Limits of Radio Telescope Arrays}
\author{Stefan J. Wijnholds~\IEEEmembership{Student member,~IEEE} and Alle-Jan
  van der Veen~\IEEEmembership{Fellow,~IEEE} % <- this % stops a space
\thanks{This work was supported by the Netherlands Foundation for Research in
  Astronomy (ASTRON) and by NWO-STW under the VICI programme (DTC.5893).}% <-
\thanks{S.J.\ Wijnholds is with ASTRON, Dwingeloo, The Netherlands. A.-J.\ van
  der Veen is with the Delft University of Technology, Delft, The
  Netherlands. Email: wijnholds@astron.nl, a.j.vanderveen@tudelft.nl}}

\def\revdate{version 11 June 2008}
\def\diag{\mathrm{diag}}
\def\trace{\mathrm{tr}}
\def\vecdiag{\mathrm{vecdiag}}
\def\jcmplx{\mathrm{j}}
\def\e{\mathrm{e}}
\def\expect{\mathcal{E}}
\def\vec{\mathrm{vec}}
\def\bgamma{{\mbox{\boldmath{$\gamma$}}}}
\def\bGamma{{\mbox{\boldmath{$\Gamma$}}}}
\def\btheta{{\mbox{\boldmath{$\theta$}}}}
\def\blambda{{\mbox{\boldmath{$\lambda$}}}}
\def\bSigma{{\mbox{\boldmath{$\Sigma$}}}}
\def\bsigma{{\mbox{\boldmath{$\sigma$}}}}
\def\bPhi{{\mbox{\boldmath{$\Phi$}}}}
\def\bphi{{\mbox{\boldmath{$\phi$}}}}
\newcommand{\bcL}{\boldmath{\ensuremath{\mathcal{L}}}}
\newcommand{\bcR}{\boldmath{\ensuremath{\mathcal{R}}}}

\begin{document}

\maketitle

\begin{abstract}
  The fidelity of radio astronomical images is generally assessed by practical
  experience, i.e. using rules of thumb, although some aspects and cases have
  been treated rigorously. In this paper we present a mathematical framework
  capable of describing the fundamental limits of radio astronomical imaging
  problems. Although the data model assumes a single snapshot observation,
  i.e. variations in time and frequency are not considered, this framework is
  sufficiently general to allow extension to synthesis observations. Using
  tools from statistical signal processing and linear algebra, we discuss the
  tractability of the imaging and deconvolution problem, the redistribution of
  noise in the map by the imaging and deconvolution process, the covariance of
  the image values due to propagation of calibration errors and thermal noise
  and the upper limit on the number of sources tractable by self
  calibration. The combination of covariance of the image values and the
  number of tractable sources determines the effective noise floor achievable
  in the imaging process. The effective noise provides a better figure of
  merit than dynamic range since it includes the spatial variations of the
  noise. Our results provide handles for improving the imaging performance by
  design of the array.
\end{abstract}


\IEEEpeerreviewmaketitle

\section{Introduction}

The radio astronomical community is currently building or developing a number
of new instruments such as the low frequency array (LOFAR)
\cite{Bregman2005-1}, the square kilometer array (SKA) \cite{Hall2004-1} and
the Mileura wide field array (MWA) \cite{Lonsdale2005-1}. Imaging and self
calibration of these radio telescopes will be computationally demanding tasks
due to the large number of array elements. Much research is therefore focused
on finding clever short-cuts to reduce the amount of processing required, such
as $w$-projection \cite{Cornwell2005-1} or facet imaging \cite{Perley1994-1}
and different variants of CLEAN \cite{Cornwell1999-1}. The validity and
quality of these methods is generally assessed by practical
experience. Attempts to do a rigorous analysis are done for some aspects and
cases \cite{Schwarz1978-1, Tan1986-1, Kulkarni1989-1, Wijnholds2006-1}, but
most of the time rules of thumb are used. This paper presents the first
comprehensive mathematical framework capable of describing the fundamental
limits of radio astronomical imaging problems. The data model used in this
paper applies to snapshot observations, i.e. variations in time and frequency
are not considered. However, using a multi-measurement data model such as
those in \cite{Leshem2000-1, Veen2004-1, Tol2007-1} it is straightforward to
extend the data model to synthesis observation and still apply the framework
described herein.

The resolution of the final image (or {\em map}) is normally determined by the
size and configuration of the array and the spatial taper function. Under the
assumption that the sky is mainly empty, i.e.\ that the image contains only a
few sources, maps with higher resolution than predicted by the array
configuration (superresolution) can be made using CLEAN. The Maximum Entropy
Method (MEM) \cite{Narayan1986-1} imposes a similar constraint by aiming for a
solution that is as featureless as possible. In the array processing
literature, superresolution is achieved by high resolution direction of
arrival (DOA) estimation techniques such as MUSIC \cite{Schmidt1986-1} and
weighted subspace fitting \cite{Viberg1991-1, Viberg1991-2}. In all these
approaches the goal is to disentangle the spatial response of the array and
the source structure, a process called deconvolution. In section
\ref{sec:imaging} we formulate imaging as an estimation problem, an approach
called model based imaging, and obtain an analytic expression for its least
squares solution that allows us to formulate the deconvolution problem as a
matrix inversion problem. This provides a powerful tool to assess the
tractability of the deconvolution problem and to demonstrate the impact on the
array configuration on the deconvolution problem and the redistribution of
noise in the imaging and deconvolution process.

The dynamic range of an image is generally defined as the power ratio between
the strongest and the weakest meaningful features in the map. In practice, the
limitations of an instrument are more conveniently described by the achievable
noise floor in an imaging observation since the dynamic range strongly depends
on the strength of the strongest source within the field-of-view and because
the noise varies over the map. This noise floor is a combination of
calibration errors, thermal noise and confusion noise. In this paper the term
``effective noise'' refers to the net result of these constituents in the
image plane. In section \ref{sec:effnoise} analytical expressions are derived
that describe the components of the effective noise in terms of the covariance
of the image values, a concept which we will refer to as image covariance. The
consequences of these expressions are illustrated with a few examples in
section \ref{sec:implications}. These examples suggest that the contribution
of propagated calibration errors to the image covariance is considerably
smaller than the contribution of thermal noise even if the calibration is done
on data with similar SNR. They also indicate that self calibration causes
higher covariance between source power estimates than pure imaging does.

\emph{Notation}: Overbar $\overline{(\cdot)}$ denotes complex conjugation.
The transpose operator is denoted by $^T$, the complex conjugate (Hermitian)
transpose by $^H$ and the Moore-Penrose pseudo-inverse by $^\dagger$. The
expectation operator is denoted by $\expect \{ \cdot \}$, $\odot$ is the
element-wise matrix multiplication (Hadamard product), $(\cdot)^{\odot n}$ is
used to denote the element-wise matrix exponent with exponent $n$, $\otimes$
denotes the Kronecker product and $\circ$ is used to denote the Khatri-Rao or
column-wise Kronecker product of two matrices. $\diag(\cdot)$ converts a
vector to a diagonal matrix with the elements of the vector placed on the main
diagonal, $\vec(\cdot)$ converts a matrix to a vector by stacking the columns
of the matrix and $\vecdiag(\cdot)$ converts the main diagonal of its argument
to a column vector. $\mathrm{circulant}(\cdot)$ creates a square circulant
matrix by circularly shifting the entries of its vector argument to form its
columns. $\circledast$ will be used to denote circular convolution of two
vectors, i.e., for vectors of length $n$, $(\mathbf{x} \circledast
\mathbf{y})_j = \sum_{i=0}^{n-1} x_i y_{j-i\,\mathrm{mod}\, n}$.

For matrices and vectors of compatible dimensions, we will
frequently use the following properties:
\begin{eqnarray}
   \vec(\mathbf{A} \mathbf{B} \mathbf{C}) &=& (\mathbf{C}^T \otimes \mathbf{A})
   \vec(\mathbf{B})
\label{eq:prop1}
\\
   \vec(\mathbf{A} \diag(\mathbf{b}) \mathbf{C}) &=& (\mathbf{C}^T \circ
   \mathbf{A}) \mathbf{b}
\label{eq:prop2}
\\
   (\mathbf{A} \circ \mathbf{B})^H
   (\mathbf{C} \circ \mathbf{D})
   &=&
   \mathbf{A}^H \mathbf{C} \odot \mathbf{B}^H \mathbf{D}
\label{eq:prop3}
\\
   (\mathbf{A} \otimes \mathbf{B}) (\mathbf{C} \circ \mathbf{D}) &=&
   \mathbf{A} \mathbf{C} \circ \mathbf{B} \mathbf{D}
\label{eq:prop4}
\end{eqnarray}

\section{Data model}
\label{sec:datamodel}

Consider a phased array consisting of $p$ sensors (antennas).  Denote the
baseband output signal of the $i$th array element as $x_i(t)$ and define the
array signal vector $\mathbf{x}(t) = [x_1(t), x_2(t), \cdots, x_p(t)]^T$. We
assume the presence of $q$ source signals $s_k(t)$ impinging on the array.
These are assumed to be mutually independent i.i.d.\ Gaussian signals, and are
stacked in a $q \times 1$ vector $\mathbf{s}(t)$. Likewise the sensor noise
signals $n_i(t)$ are assumed to be mutually independent Gaussian signals and
are stacked in a $p \times 1$ vector $\mathbf{n}(t)$. We assume that the
narrowband condition holds \cite{Zatman1998-1}. We can then describe, for the
$k$th source signal, the phase delay differences over the $p$ receiving
elements due to the propagation geometry by a $p$-dimensional spatial
signature vector $\mathbf{a}_k$.  The $q$ spatial signature vectors are
assumed to be known (known source locations and array geometry).

The sensors are assumed to have the same direction dependent gain behavior
which is described by gain factors $g_{0k}$ towards the $q$ source signals
received by the array. These can be collected in a matrix $\mathbf{G}_0 =
\diag([g_{01}, g_{02}, \cdots, g_{0q}])$. The direction independent gains and
phases can be described as $\bgamma = [\gamma_1, \gamma_2, \cdots,
\gamma_p]^T$ and $\bphi = [\e^{\jcmplx\phi_1}, \e^{\jcmplx\phi_2}, \cdots,
\e^{\jcmplx\phi_p}]^T$ respectively, with corresponding diagonal matrix forms
$\bGamma = \diag(\bgamma)$ and $\bPhi = \diag(\bphi)$. With these definitions,
the array signal vector can be described as
\begin{equation}
\mathbf{x}(t) = \bGamma\bPhi \left ( \sum_{k=1}^q \mathbf{a}_k g_{0k} s_k(t)
\right ) + \mathbf{n}(t) = \mathbf{G} \mathbf{A} \mathbf{G}_0 \mathbf{s}(t) +
\mathbf{n}(t)\label{eq:sigvec}
\end{equation}
where $\mathbf{A} = [\mathbf{a}_1, \cdots, \mathbf{a}_k]$ (size $p \times q$)
and $\mathbf{G} = \bGamma \bPhi$.

The signal is sampled with period $T$ and $N$ sample vectors are stacked into
a data matrix $\mathbf{X} = [\mathbf{x}(T), \mathbf{x}(2T), \cdots,
\mathbf{x}(NT)]$. The covariance matrix of $\mathbf{x}(t)$ is $\mathbf{R} =
\expect \{ \mathbf{x}(t) \mathbf{x}^H(t) \}$ and is estimated by
$\widehat{\mathbf{R}} = N^{-1} \mathbf{X} \mathbf{X}^H$. The number of samples
$N$ in a snapshot observation is equal to the product of bandwidth and
integration time and typically ranges from $10^3$ (1 s, 1 kHz) to $10^6$ (10
s, 100 kHz) in radio astronomical applications. Likewise, the source signal
covariance $\bSigma_s = \diag(\bsigma_s)$ where $\bsigma_s = [\sigma_{s1}^2,
\sigma_{s2}^2, \cdots, \sigma_{sq}^2]^T$ and the noise covariance matrix is
$\bSigma_n = \diag(\bsigma_n)$ where $\bsigma_n = [\sigma_{n1}^2,
\sigma_{n2}^2, \cdots, \sigma_{np}^2]^T$. Then the model for the covariance
matrix for a snapshot observation $\mathbf{R}$ based on \eqref{eq:sigvec} is
\begin{equation}
\mathbf{R} = \mathbf{G} \mathbf{A} \mathbf{G}_0 \bSigma_s \mathbf{G}_0^H
\mathbf{A}^H \mathbf{G}^H + \bSigma_n.\label{eq:R}
\end{equation}

If the directional response of the antennas is known, $\mathbf{G}_0$ can be
absorbed in $\mathbf{A}$. If $\mathbf{G}_0$ and $\bSigma_s$ are both unknown,
we can introduce
\begin{eqnarray}
\bSigma & = & \mathbf{G}_0 \bSigma_s \mathbf{G_0}^H \nonumber\\
& = & \diag([ |g_{01}|^2\sigma_{s1}, \cdots, |g_{0q}|^2\sigma_{sq}]) =
\diag(\bsigma)
\end{eqnarray}
with real valued elements $\bsigma = [\sigma_1^2, \sigma_2^2, \cdots,
\sigma_q^2]^T$. We may then restate \eqref{eq:R} as
\begin{equation}
\mathbf{R} = \mathbf{G} \mathbf{A} \bSigma \mathbf{A}^H \mathbf{G}^H +
\bSigma_n. \label{eq:R2}
\end{equation}

The $i$th element of the sensor array is located at $\mathbf{r}_i = [x_i, y_i,
z_i]^T$. These positions can be stacked in a matrix $\bcR = [\mathbf{r}_1,
\mathbf{r}_2, \cdots, \mathbf{r}_p]^T$ (size $p \times 3$). The position of
the $k$th source can be denoted by the unit vector $\mathbf{l}_k = [l_k, m_k,
n_k]^T$. The source positions can be stacked in a matrix $\bcL =
[\mathbf{l}_1, \mathbf{l}_2, \cdots, \mathbf{l}_q]^T$ (size $q \times 3$). The
spatial signature matrix $\mathbf{A}$ can thus be described by
\begin{equation}
\mathbf{A} = \exp \left ( -\jcmplx \frac{2\pi}{\lambda} \bcR \bcL^T \right
) \label{eq:A}
\end{equation}
where the exponential function is applied element-wise to its argument. In the
remainder of this paper we will specialize to a planar array having $z_i = 0$
for convenience of presentation but without loss of generality.

\section{Imaging and deconvolution}
\label{sec:imaging}

\subsection{Beam forming versus model based imaging}

The imaging process transforms the covariances of the received signals (called
{\em visibilities} in radio astronomy) to an image of the source structure
within the field-of-view of the receivers. In array processing terms, it can
be described as follows \cite{Leshem2000-1}. To determine the power of a
signal received from a particular direction $(l, m, n)$, a weight vector
\begin{eqnarray}
\mathbf{w} & = & (\mathbf{a}^\dagger)^H = \exp \left ( - \jcmplx
  \frac{2\pi}{\lambda} \mathcal{R} [l, m, n]^T \right )^{\dagger H} 
  \nonumber\\
& = & \frac{1}{p} \exp \left (-\jcmplx \frac{2\pi}{\lambda} \mathcal{R} [l, m,
  n]^T \right ) \label{eq:BFweight}
\end{eqnarray}
is assigned to the array signal vector $\mathbf{x}(t)$. The operation $y(t) =
\mathbf{w}^H \mathbf{x}(t)$ is generally called
beamforming and can be regarded as a spatially matched filter. Equation
(\ref{eq:BFweight}) represents the most basic beamformer that assumes the
presence of only a single source and only corrects the
signal delays due to the array geometry. These weights can be adapted to
correct the complex gain differences between the receiving elements
$\mathbf{G}$ derived from calibration measurements \cite{Wijnholds2006-3},
nulling of interfering sources \cite{Veen2004-1} and spatial tapering of the
array \cite{Wijnholds2006-4}.

\begin{figure}
\centering
\includegraphics[width=8.5cm]{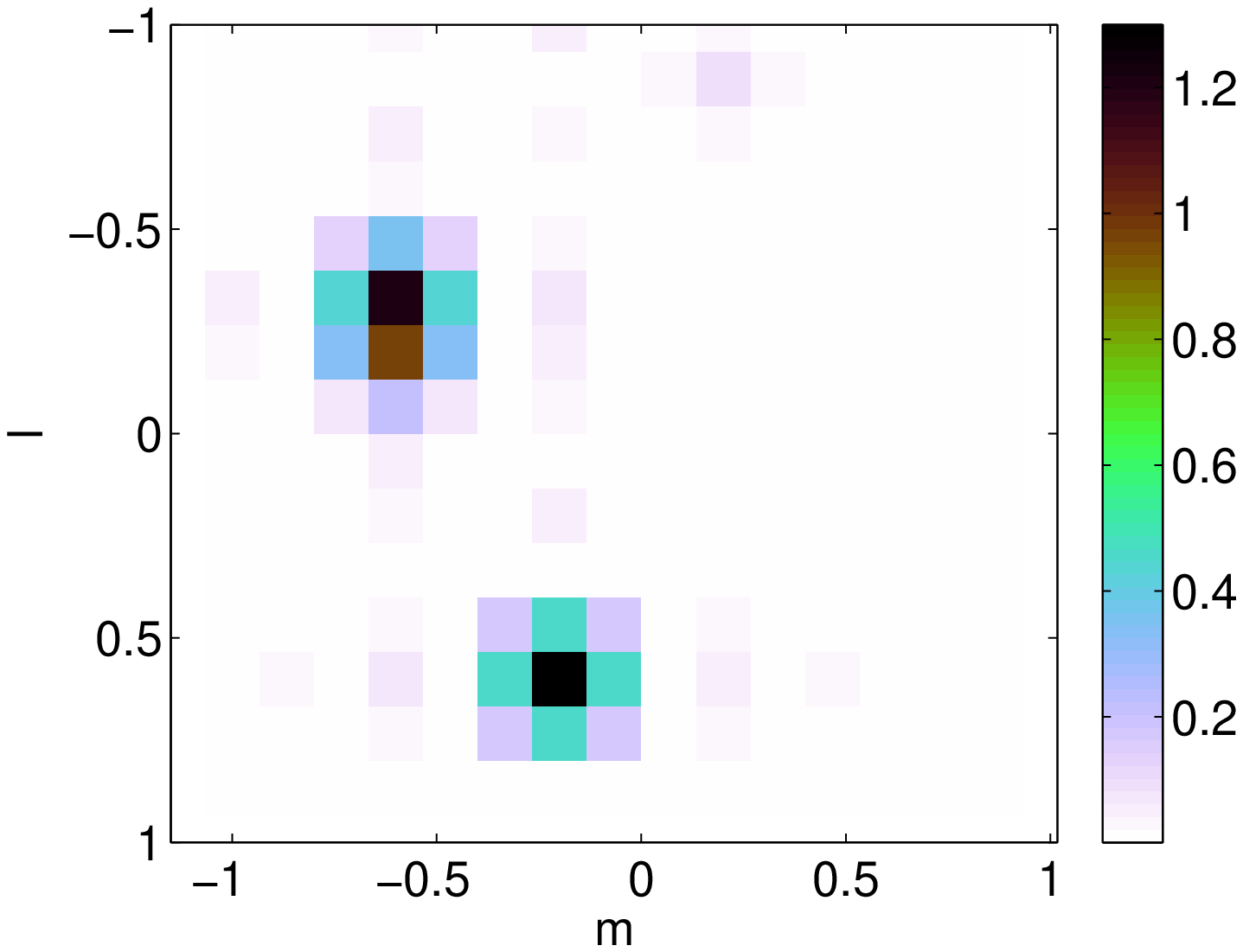}
\includegraphics[width=8.5cm]{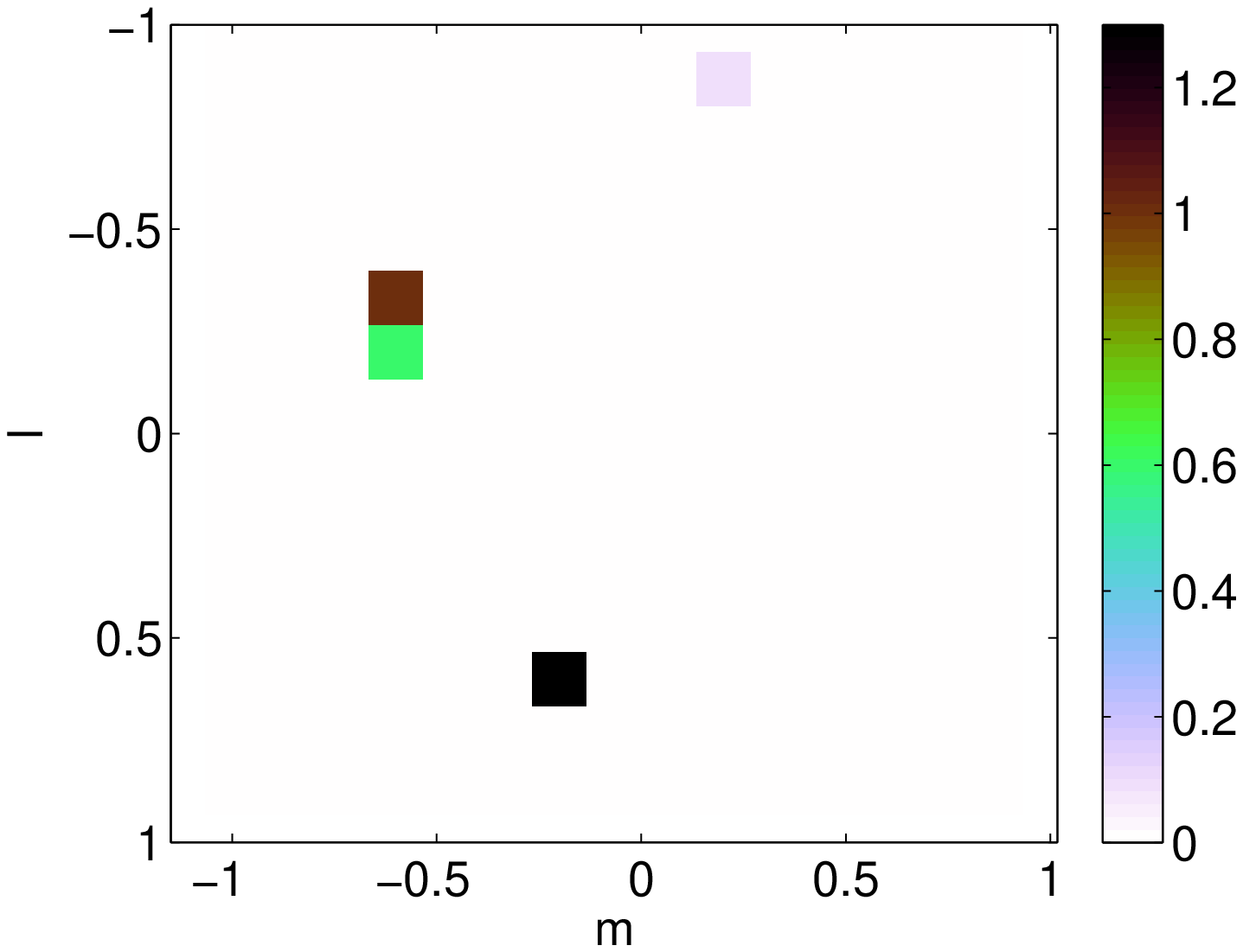}
\caption{ ({\it a}) Image obtained by normal imaging without deconvolution
  as in (\ref{eq:imaging}), showing the sources and their side lobe patterns.
  ({\it b}) Image obtained by model based imaging as in
  (\ref{eq:modelbasedimaging}), which estimates the power at every pixel
  simultaneously, resulting in a deconvolved image showing only the sources
  without the array response.}
\label{fig:imagingdemo}
\end{figure}

The image value at $(l,m, n)$ is equal to the expected output power of
the beamformer when pointed into that direction, and can be computed
directly from the array covariance matrix $\widehat{\mathbf{R}}$ as
\begin{equation}
\widehat{i} \left ( l, m, n \right ) = \mathbf{w}^H \widehat{\mathbf{R}}
\mathbf{w}. \label{eq:imaging}
\end{equation}
For weights defined as in (\ref{eq:BFweight}), this is known as
{\em direct Fourier transform imaging}.
To create an image, $\mathbf{w}$ is scanned over all relevant $(l,m,n)$.
The required weights can be stacked into a single matrix $\mathbf{W}$.
Since
$\mathbf{w}^H \widehat{\mathbf{R}} \mathbf{w} = ( \overline{\mathbf{w}} \otimes
\mathbf{w} )^H \vec ( \widehat{\mathbf{R}} )$,
we can stack all image values in a single vector $\widehat{\mathbf{i}}$
and write
\begin{equation}
\widehat{\mathbf{i}} = 
%%% \frac{1}{p^2} ( \overline{\mathbf{A}} \circ \mathbf{A} )^H 
(\overline{\mathbf{W}} \circ \mathbf{W} )^H 
\vec ( \widehat{\mathbf{R}} ). 
\label{eq:imaging1}
\end{equation}
If we only want to image at the source locations, we have $\mathbf{W} =
\frac{1}{p} \mathbf{A}$.  A typical model assumption is that there is a source
present at every pixel location, in which case
\begin{equation}
\widehat{\mathbf{i}}_{BF} = 
\frac{1}{p^2} ( \overline{\mathbf{A}} \circ \mathbf{A} )^H 
\vec ( \widehat{\mathbf{R}} ). 
\label{eq:BFimaging}
\end{equation}
This is the classical {\em dirty image}.  

Let us assume momentarily that $\mathbf{G}= \mathbf{I}$ and $\mathbf{\Sigma}_n
= 0$. Inserting the data model (\ref{eq:R}), or $\vec(\mathbf{R}) =
(\overline{\mathbf{A}} \circ \mathbf{A}) \bsigma$, into
(\ref{eq:BFimaging}) gives
\begin{eqnarray}
\mathbf{i}_{BF} & = & \expect\{\widehat{\mathbf{i}}_{BF}\} =  \frac{1}{p^2} (
\overline{\mathbf{A}} \circ \mathbf{A} )^H (\overline{\mathbf{A}} \circ
\mathbf{A}) \bsigma \nonumber\\
& = & \frac{1}{p^2} ( \overline{\mathbf{A}}^H \overline{\mathbf{A}} \odot 
\mathbf{A}^H \mathbf{A}) \bsigma
\label{eq:BFimaging1}
\end{eqnarray}
This shows that the dirty image is not equal to the true source structure. To
understand the physical meaning of this term, consider the product
$\mathbf{a}_i^H \mathbf{a}_j$, where the indices $i$ and $j$ refer to the
respective columns of $\mathbf{A}$. Using \eqref{eq:A} this can be written
explicitly as
\begin{eqnarray}
\mathbf{a}_i^H \mathbf{a}_j & = & \exp \left ( -\jcmplx \frac{2\pi}{\lambda}
  \bcR \mathbf{l}_i \right )^H \exp \left ( -\jcmplx \frac{2\pi}{\lambda} \bcR
  \mathbf{l}_j \right ) \nonumber\\
& = & \sum_{n = 1}^p \exp \left ( \jcmplx \frac{2\pi}{\lambda} \mathbf{r}_n^T
  \left ( \mathbf{l}_i - \mathbf{l}_j \right ) \right ).
\end{eqnarray}
The physical interpretation of the inner product between the two spatial
signature vectors is that it measures the sensitivity of the array to signals 
coming from
direction $\mathbf{l}_j$ while the array is steered towards $\mathbf{l}_i$. The
product $\mathbf{A}^H \mathbf{a}_j$ thus describes the array sensitivity for
all directions of interest stacked in $\bcL$ when pointed to
$\mathbf{l}_j$. It therefore provides the
array voltage response or array voltage beam pattern centered around
$\mathbf{l}_j$, 
\begin{equation}
\mathbf{b}_V \left ( \mathbf{l}_j \right ) = \mathbf{A}^H \mathbf{a}_j.
\end{equation}
With $\mathbf{A}$ defined as in \eqref{eq:A}, this shows that the voltage
beam pattern is just the Fourier transform of the spatial weighting function
resulting from the array configuration and the weighting of the array
elements. The corresponding power beam pattern can be calculated as
\begin{equation}
\mathbf{b}_P \left ( \mathbf{l}_j \right ) = \overline{\mathbf{b}}_V \left (
  \mathbf{l}_j \right ) \odot \mathbf{b}_V \left ( \mathbf{l}_j \right ) =
\overline{\mathbf{A}}^H \overline{\mathbf{a}}_j \odot \mathbf{A}^H
\mathbf{a}_j.
\end{equation}
The factor $\overline{\mathbf{A}}^H \overline{\mathbf{A}} \odot
\mathbf{A}^H \mathbf{A}$ in (\ref{eq:BFimaging1}) 
can thus be interpreted as a convolution by the
Fourier transform of the spatial distribution of baseline vectors, which is
known as the array beam pattern or {\em dirty beam} \cite{Thompson2001-1}.

This effect is illustrated in Fig.\ \ref{fig:imagingdemo}($a$). This image is
the result of a simulated observation with an $8 \times 8$ half wavelength
spaced (i.e., spatially Nyquist sampled) 2D uniform rectangular array
(URA). The grid of image values on the sky is taken such that the first
Nyquist zone is appropriately sampled. The underlying source model contains
four sources at grid points $(-0.33, -0.6, 0.73)$, $(-0.2, -0.6, 0.77)$,
$(0.6, -0.2, 0.77)$ and $(0.87, 0.2, 0.46)$ respectively and $\bsigma = [1,
0.6, 1.3, 0.1]^T$. This source and array configuration will be used throughout
this paper unless stated otherwise. The map in Fig.\
\ref{fig:imagingdemo}($a$) clearly shows these four (or three, if one regards
the two sources on neighboring grid points as a single extended source) being
convolved with the array beam pattern.

Following a model based approach, the deconvolution problem can be formulated
as a maximum likelihood (ML) estimation problem, that should provide a
statistically efficient estimate of the parameters. Since all signals are
assumed to be i.i.d. Gaussian signals, the derivation is standard and the ML
estimates are obtained by minimizing the negative log-likelihood function
\cite{Ottersten1998-1}
\begin{equation}
\widehat{\bsigma} = \underset{\bsigma}{\mathrm{argmin}} \left ( \ln \left |
    \mathbf{R} (\bsigma) \right | + \trace \left ( \mathbf{R}^{-1} ( \bsigma )
    \widehat{\mathbf{R}} \right ) \right ).
\end{equation}
It does not seem possible to solve this minimization problem is closed
form, but a weighted least squares covariance matching approach is known to
lead to estimates that are, for a large number of samples, equivalent to ML
estimates and therefore asymptotically efficient \cite{Ottersten1998-1}. The
problem can thus be reformulated as
\begin{eqnarray}
\widehat{\bsigma} & = & \underset{\bsigma}{\mathrm{argmin}} \Big \Arrowvert
\mathbf{W}_c \left ( \widehat{\mathbf{R}} - \bSigma_n \right ) \mathbf{W}_c -
\nonumber\\
&  & \mathbf{W}_c \mathbf{G} \mathbf{A} \bSigma \mathbf{A}^H \mathbf{G}^H
\mathbf{W}_c \Big \Arrowvert_F^2 \nonumber\\
& = & \underset{\bsigma}{\mathrm{argmin}} \Big \Arrowvert \left (
  \overline{\mathbf{W}}_c \otimes \mathbf{W}_c \right ) \vec \left (
  \widehat{\mathbf{R}} - \bSigma_n \right ) - \nonumber\\
& & \left ( \overline{\mathbf{W}_c \mathbf{G} \mathbf{A}} \right ) \circ \left
  ( \mathbf{W}_c \mathbf{G} \mathbf{A} \right ) \bsigma_n \Big \Arrowvert_F^2
\label{eq:sigmalsq}
\end{eqnarray}

The solution is given by
\begin{equation}
\widehat\bsigma = \left ( \left ( \overline{\mathbf{W}_c \mathbf{G}
      \mathbf{A}} \right ) \circ \left ( \mathbf{W}_c \mathbf{G} \mathbf{A}
  \right ) \right )^\dagger \left ( \overline{\mathbf{W}}_c \otimes
  \mathbf{W}_c \right ) \vec \left ( \widehat{\mathbf{R}} - \bSigma_n \right
) \label{eq:pre-srcpowerest}
\end{equation}
Optimal weighting is provided by $\mathbf{W}_c =
\mathbf{R}^{-\frac{1}{2}}$. Since radio astronomical sources are generally
very weak with the strongest source in the field having an instantaneous SNR
in the order of 0.01, we can introduce the approximation $\mathbf{R} \approx
\sigma_n^2 \mathbf{I}$ for an array of identical elements for convenience of
notation. This reduces \eqref{eq:pre-srcpowerest} to
\begin{eqnarray}
\widehat{\bsigma} & = & \left ( \overline{\mathbf{GA}} \circ \mathbf{GA}
\right )^\dagger \vec \left ( \widehat{\mathbf{R}} - \bSigma_n \right
). \label{eq:srcpowerest}
\end{eqnarray}

One may argue that this requires one to know where the sources are before
doing the imaging. This is generally solved by simultaneously estimating the
source locations and source powers. Although the CLEAN algorithm has not yet
been fully analyzed, it can be regarded as an iterative procedure to do this
\cite{Leshem2000-1}. It is instructive, however, to use \eqref{eq:srcpowerest}
for imaging by estimating the power on every image point (pixel), i.e., by
assuming a data model with a source present at every pixel.  We can simplify
(\ref{eq:srcpowerest}) by replacing the Moore-Penrose pseudo-inverse by the
left pseudo-inverse, to obtain the image vector
\begin{eqnarray}
\widehat{\mathbf{i}} & = & \left ( \left ( \overline{\mathbf{GA}} \circ
    \mathbf{GA} \right )^H \left ( \overline{\mathbf{GA}} \circ \mathbf{GA}
  \right ) \right )^{-1} \times \nonumber\\
& & \times \left ( \overline{\mathbf{GA}} \circ \mathbf{GA} \right )^H \vec
\left ( \widehat{\mathbf{R}} - \bSigma_n \right ) \nonumber\\
& = & \left ( \overline{\mathbf{A}}^H \bGamma^2 \overline{\mathbf{A}} \odot
    \mathbf{A}^H \bGamma^2 \mathbf{A} \right )^{-1} \times \nonumber\\
& & \times \left ( \overline{\mathbf{GA}} \circ \mathbf{GA} \right )^H \vec
\left ( \widehat{\mathbf{R}} - \bSigma_n \right
). \label{eq:modelbasedimaging}
\end{eqnarray}

The first factor in this equation represents the deconvolution operation. It
is therefore convenient to introduce the deconvolution matrix $\mathbf{M} =
\overline{\mathbf{A}}^H \bGamma^2 \overline{\mathbf{A}} \odot \mathbf{A}^H
\bGamma^2 \mathbf{A} = \left | \mathbf{A}^H \bGamma^2 \mathbf{A} \right
|^{\odot 2}$.  This provides a powerful check on the sampling of the image
plane. If the image plane is oversampled, i.e., if too many image points are
defined, this matrix will be singular. This property demonstrates that high
resolution imaging is only possible if a limited number of sources is present,
i.e., if the number of sources is much smaller than the number of resolution
elements in the field-of-view. The condition number of the deconvolution
matrix, which provides a measure on the magnification of measurement noise, is
discussed in more detail in Sec. \ref{subsec:cond_number}. This mostly empty
field-of-view is commonly assumed in astronomical imaging and this assumption
is one of the reasons why CLEAN and MEM work in practice. Fig.\
\ref{fig:imagingdemo}($b$) shows the image obtained by applying
\eqref{eq:modelbasedimaging} to the $8\times 8$ URA. Comparison with the image
obtained using \eqref{eq:BFimaging} clearly shows the effectiveness of the
model based imaging approach in suppressing the array beam pattern.

\subsection{Noise redistribution}
\label{ssec:noiseredistribution}

If imaging is done without deconvolution by using \eqref{eq:BFimaging},
the thermal noise adds a constant value to all image values. This can be
illustrated by assuming that $\expect \left \{ \widehat{\mathbf{R}} \right \}
= \bSigma_n$, i.e. by assuming that the image is completely dominated by
thermal noise. The expected value of the image then becomes
\begin{eqnarray}
\mathbf{i}_{BF} & = & \frac{1}{p^2} \left ( \overline{\mathbf{A}} \circ
  \mathbf{A} \right )^H \vec \left ( \bSigma_n \right ) \nonumber\\
& = & \frac{1}{p^2} \left ( \overline{\mathbf{A}} \odot \mathbf{A} \right )^H
\bsigma_n \nonumber\\
& = & \frac{\mathbf{1}^T \bsigma_n}{p^2} \mathbf{1}
\end{eqnarray}
where we used the fact that all elements of $\mathbf{A}$ have unit
amplitude. This equation describes an image where all values are equal to the
average thermal noise per baseline.

If the imaging process involves deconvolution, the result is described by
\eqref{eq:modelbasedimaging}. For simplicity we will assume that we have an
array of identical elements, so that we can set $\mathbf{G} = \mathbf{I}$.
Further, to illustrate the effect, we momentarily omit the correction by
$\mathbf{\Sigma}_n$ in \eqref{eq:modelbasedimaging}. In this case, the
expected value of the image is
\begin{eqnarray}
\mathbf{i} & = & \left ( \overline{\mathbf{A}}^H \overline{\mathbf{A}} \odot
  \mathbf{A}^H \mathbf{A} \right )^{-1} \left ( \overline{\mathbf{A}} \circ
  \mathbf{A} \right )^H \bSigma_n \nonumber\\
& = & \left ( \left | \mathbf{A}^H \mathbf{A} \right |^{\odot 2} \right )^{-1}
\left ( \overline{\mathbf{A}} \odot \mathbf{A} \right )^H \bsigma_n
\nonumber\\
& = & \left ( \left | \mathbf{A}^H \mathbf{A} \right |^{\odot 2} \right )^{-1}
\left ( \mathbf{1}^T \bsigma_n \right ) \mathbf{1}.
\end{eqnarray}
In this case, the homogeneity of the thermal noise distribution in the map
depends on the row sums of $\left ( \left | \mathbf{A}^H \mathbf{A} \right
  |^{\odot 2} \right )^{-1}$ being constant. If this is true, the model based
image using \eqref{eq:modelbasedimaging} is analogous to the beamformed image
based on \eqref{eq:BFimaging}. A special case is the situation in which the
columns of $\mathbf{A}$ are orthonormal.

\begin{figure}
\centering
\includegraphics[width=8.5cm]{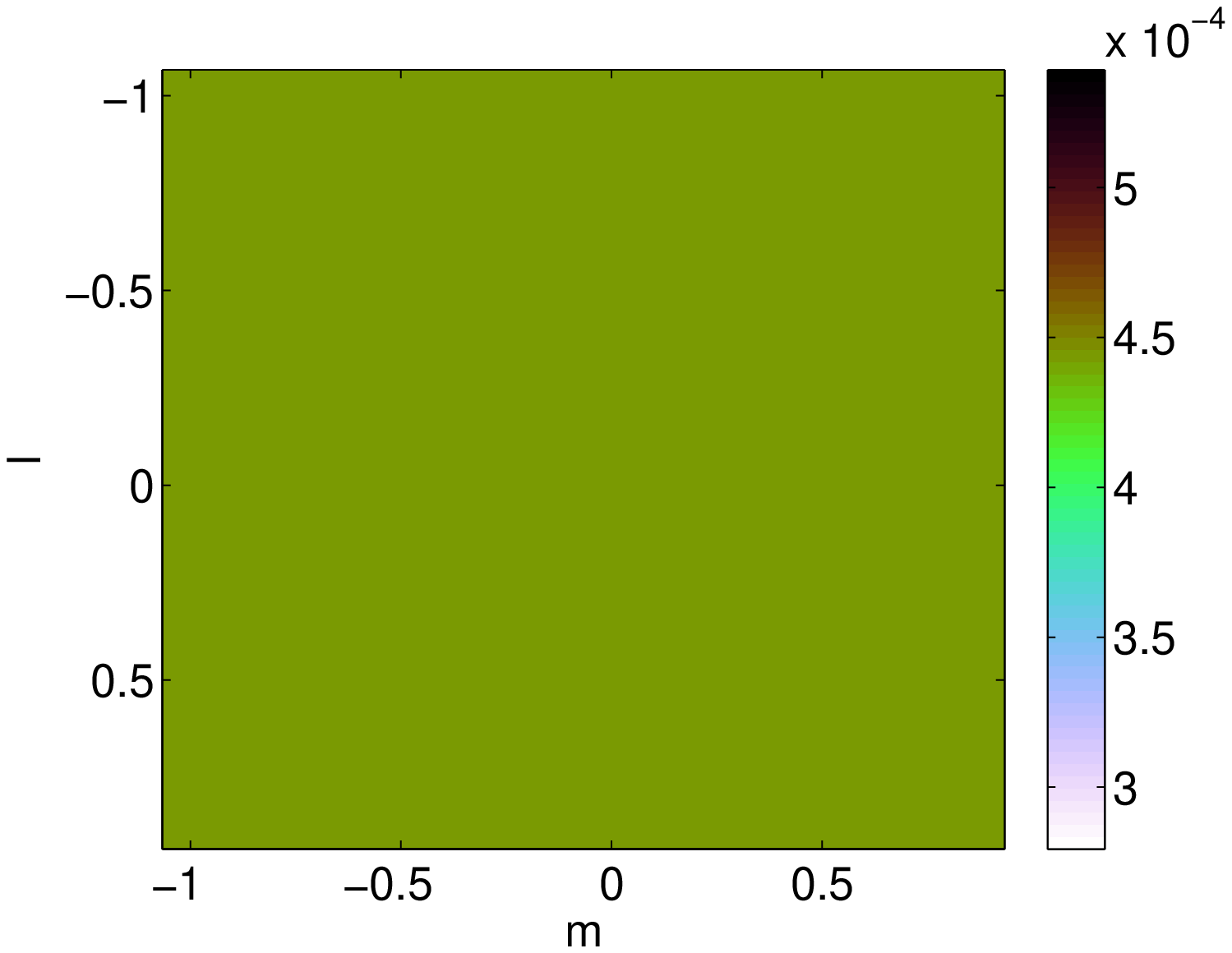}
\includegraphics[width=8.5cm]{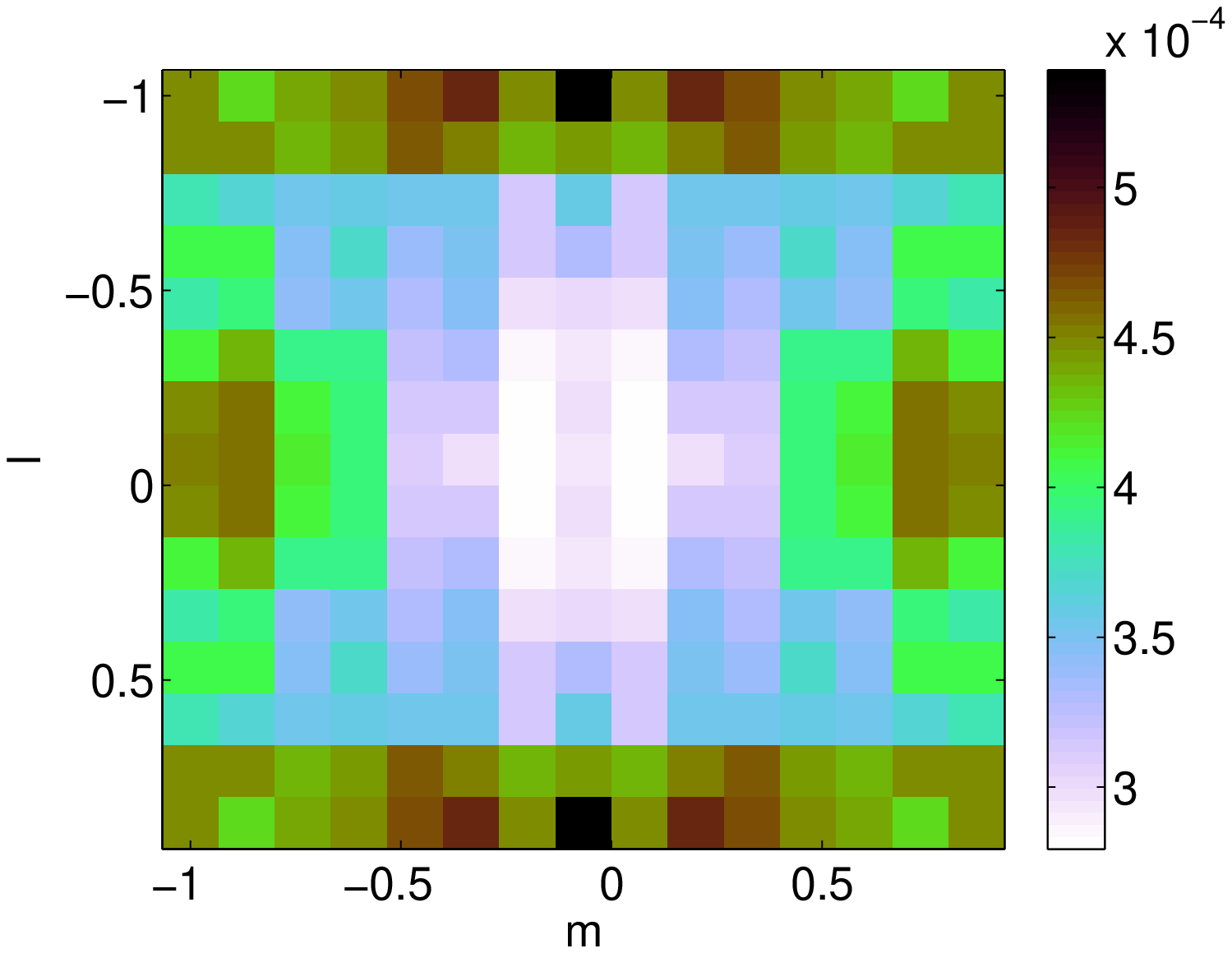}
\caption{($a$) Imaging with deconvolution using
  an $8 \times 8$ half wavelength spaced array for a Nyquist sampled image
  assuming $\mathbf{R} = 0.1 \mathbf{I}$ (an empty field with only thermal
  noise). ($b$) Imaging
  result for a five armed array, each arm being an eight element half
  wavelength spaced ULA. \label{fig:redistribution}}
\end{figure}

Otherwise, the structure is more complicated. This is illustrated in Fig.\
\ref{fig:redistribution} which compares the noise distribution in the image
plane of the $8\times 8$ URA by assuming $\mathbf{R} = 0.1 \mathbf{I}$ with
the corresponding image for a five armed array, each arm being an eight
element half wavelength spaced Uniform Linear Array (ULA).  The impact of the
redistribution of noise can be reduced by estimating the receiver noise powers
and subtracting these estimates from the array covariance matrix as described
by \eqref{eq:modelbasedimaging}. In most astronomical imaging algorithms, the
autocorrelations are generally ignored completely thus effectively introducing
a small negative system noise since the autocorrelations represent the power
sum of the source signals and the noise.

\subsection{Deconvolution matrix condition number}
\label{subsec:cond_number}

The deconvolution matrix $\mathbf{M}$ not only causes a redistribution of
noise over the map, but also determines whether the deconvolution is a well
conditioned problem. If the deconvolution matrix is not invertible, the
problem is ill-posed and additional constraints are required to obtain a
unique solution. Different choices for these constraints or even the rigor
with which they are applied, lead to different imaging results for CLEAN and
MEM based on the same data. In some cases, this may even lead to different
interpretation of the final maps \cite{Narayan1986-1}. These problems arise
due to over-interpretation of the data by allowing for more image points
(parameters) than can be justified by the data. In these situations, the
condition number of the deconvolution matrix will be infinitely large. Even if
the deconvolution matrix is invertible, its condition number may be
unacceptably high in view of the SNR of the data: the condition number is a
measure for the magnification of measurement noise \cite{Golub1984-1}. The
condition number thus provides a powerful diagnostic tool to assess the
feasibility of the deconvolution problem at hand.

It is instructive to analyze a half wavelength spaced 1D ULA with identical
elements, i.e. with $\mathbf{G} = \mathbf{I}$, sampling the sky on a regular
grid. In this case $\mathbf{A}$ represents a Fourier transform 
mapping the
spatial frequencies on the sky to the spatial samples describing the
electromagnetic field over the array aperture. As demonstrated in the previous
section, these spatial frequencies will be convolved in the imaging process
with the Fourier transform of the array aperture taper or voltage beam
pattern, which can be easily calculated for $\mathbf{l} = \mathbf{0}$:
\begin{equation}
\mathbf{b}_V \left ( \mathbf{0} \right ) 
= \mathbf{A}^H \mathbf{a}(\mathbf{0})
= \mathcal{FT} 
    \left ( 
    \left[ \begin{array}{@{}c@{}} 
	\mathbf{1}_p \\ \mathbf{0}_{n-p} 
    \end{array}\right] 
    \right ).
\end{equation}
Here $\cal{FT}$ denotes the Fourier transform, $n$ is the total number of
image points, $p$ is the number of elements in the array and $\mathbf{0}_p$
and $\mathbf{1}_p$ denote $p \times 1$ vectors containing zeros and ones
respectively. The corresponding power beam pattern is
\begin{eqnarray}
  \mathbf{b}_P \left ( \mathbf{0} \right ) & = &  \overline{\mathbf{b}}_V \left
    ( \mathbf{0} \right ) \odot \mathbf{b}_V \left ( \mathbf{0} \right
  )\nonumber\\
  & = & \mathcal{FT} \left ( \left[ \begin{array}{@{}c@{}} \mathbf{1}_p \\
        \mathbf{0}_{n-p} \end{array}\right]  \circledast
    \left[ \begin{array}{@{}c@{}}  \mathbf{1}_p \\
        \mathbf{0}_{n-p} \end{array}\right]  \right ).
\end{eqnarray}
If the columns of $\mathbf{A}$ are ordered such that they describe the array
response vectors for the regularly spaced DOAs starting with $\mathbf{l}_1 =
\mathbf{0}$, it is easily seen that
\begin{equation}
  \mathbf{M} = \overline{\mathbf{A}}^H \overline{\mathbf{A}} \odot \mathbf{A}^H
  \mathbf{A} = \mathrm{circulant} \left ( \mathbf{b}_P \left ( \mathbf{0} \right
    ) \right ),
\end{equation}
i.e., that the deconvolution matrix for a 1D ULA equidistantly sampling the
image plane is a circulant matrix. Since $\mathbf{M}$ is a circulant matrix,
its eigenvalues $\blambda = [\lambda_1, \lambda_2, \cdots, \lambda_n]^T$ are
given by the Fourier transform of $\mathbf{b}_P (\mathbf{0})$ 
\cite{Moon2000-1}, or
\begin{eqnarray}
\blambda & = & \mathcal{FT} \left ( \mathbf{b}_P (\mathbf{0})\right ) 
\nonumber\\
& = & \mathcal{FT} \left ( \mathcal{FT} \left ( 
    \left[ \begin{array}{@{}c@{}} \mathbf{1}_p \\ \mathbf{0}_{n-p} \end{array}\right] 
    \circledast
    \left[ \begin{array}{@{}c@{}} \mathbf{1}_p \\ \mathbf{0}_{n-p} \end{array}\right] 
  \right ) \right ) \nonumber\\
& = & 
    \left[ \begin{array}{@{}c@{}} \mathbf{1}_p \\ \mathbf{0}_{n-p} \end{array}\right] 
    \circledast
    \left[ \begin{array}{@{}c@{}} \mathbf{1}_p \\ \mathbf{0}_{n-p} \end{array}\right] 
\end{eqnarray}
since $\mathcal{FT} ( \cdot ) = \mathcal{FT}^{-1} ( \cdot )$ for real
symmetric functions.

For Hermitian matrices, the condition number $\kappa$ is given by the ratio
of the largest and smallest eigenvalue, i.e. $\kappa = \lambda_{max} /
\lambda_{min}$ \cite{Moon2000-1}.
If the image plane is Nyquist
sampled, $n = 2p-1$ and
\begin{equation}
\blambda = \left [ 1, 2, \cdots, p-1, p, p-1, \cdots, 2, 1 \right ]^T.
\end{equation}
In this case the condition number of $\mathbf{M}$ is
\begin{equation}
\kappa = \frac{\lambda_{max}}{\lambda_{min}} = \frac{p}{1} = p,
\end{equation}
thus $\mathbf{M}$ is invertible.  The deconvolution
problem is therefore well-posed and has a unique solution.

If the image plane is undersampled with $n < 2p-1$ samples, then
\begin{equation}
\blambda = \left [ p-\frac{n-1}{2}, \cdots, p-1, p, p-1, \cdots,
  p-\frac{n-1}{2} \right ]^T
\end{equation}
and $\kappa = \frac{2p}{2p - (n -1)}$. The deconvolution problem in itself is
thus well-posed and has a unique solution. However, from Fourier theory we
know that aliasing effects may occur due to undersampling.

If the image plane is oversampled with $n > 2p-1$ samples,  then
\begin{equation}
\blambda = \left [ \cdots, 0, 1, 2, \cdots, p-1, p, p-1, \cdots, 2, 1, 0,
  \cdots \right ]^T
\end{equation}
and $\kappa = \infty$. In this case, the deconvolution problem is 
ill-posed and thus not solvable without introducing additional boundary
conditions to constrain the problem.

\begin{figure}
\centering
\includegraphics[width=8.5cm]{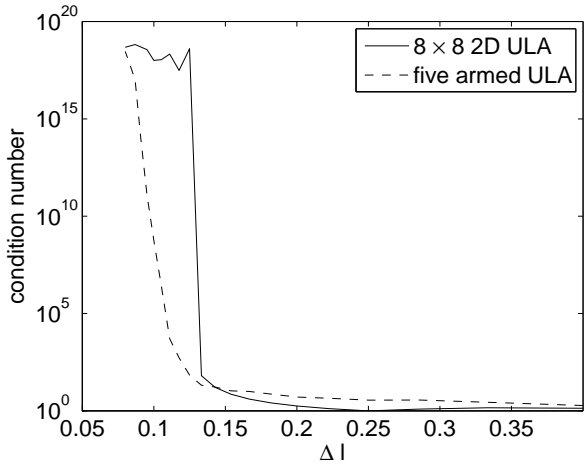}
\caption{This plot shows the condition number of the deconvolution matrix as
  function of the image resolution for the $8 \times 8$ half wavelength spaced
  array and the five armed array with each arm being an eight element half
  wavelength spaced ULA. \label{fig:conditionnumber}}
\end{figure}

This analysis shows that, for a 1D ULA, the condition number slowly increases
up to Nyquist sampling of the image plane and then jumps to infinity. Since a
URA is just the 2D analog of a 1D ULA, this behavior is also expected for the
$8\times 8$ URA introduced earlier. This conjecture is confirmed in Fig.\
\ref{fig:conditionnumber} which shows the condition number of the
deconvolution matrix as function of image resolution. This figure also shows
the corresponding curve for the five armed array introduced earlier to
demonstrate the impact of less regular and sparser sampling of the array
aperture. Although the array diameter is nearly twice as large, it does not
provide twice the resolution due to sparser sampling of the aperture
plane. This plot also demonstrates that a less regular array also may have a
less strict cut-off: the transition of the condition number from small values
to infinity is a gradual one. For array processing problems, this means that
the user should decide which value of the condition number (or noise
enhancement) is still acceptable.

Regularization is commonly used to avoid uninvertability of matrices. In radio
astronomical imaging where most sources have a low SNR, this would lead to
imperfect deconvolution causing the weakest sources in the field to be drowned
in the imperfectly removed array response pattern of the strongest
sources. However, several forms of implicit regularization have been studied
to handle special cases like strong interference \cite{Tol2005-1}.

\section{Effective noise}
\label{sec:effnoise}

Equation \eqref{eq:modelbasedimaging} shows that calibration and imaging are
strongly coupled. Knowledge of the instrumental parameters is required to
obtain the proper image. People have approached this problem in two ways. In
the first approach calibration and imaging are treated as separate steps,
i.e., the instrumental parameters are estimated first by a calibration
measurement and consecutively applied to the actual measurement data. The
second approach is self calibration which regards the estimation of
instrumental and image parameters as a single parameter estimation problem
\cite{Cornwell1994-1, Flanagan1999-1, Pearson1984-1, Tol2007-1}.

In either case the achievable dynamic range is limited by the combination of
estimation errors, thermal noise and confusion noise. Together, they determine
the effective noise in the image which need not be homogeneous over the field
of interest. In this section a number of analytical expressions are derived
that describe these contributions in terms of the data model presented in
Section \ref{sec:datamodel}. The implications will be discussed in 
Section \ref{sec:implications}.

\subsection{Noise in self calibrated images}
\label{sec:CRB}

In self calibration the instrumental and image parameters are estimated
simultaneously.  Self calibration based on the data model presented above can
thus be described as simultaneous estimation of the omni-directional complex
gains, the apparent source powers, the source locations and the receiver noise
powers, i.e., of a parameter vector
\begin{eqnarray}
\btheta & =& [\gamma_1, \cdots, \gamma_p, \phi_2, \cdots, \phi_p, \sigma_2^2,
\cdots, \sigma_q^2, \sigma_{n1}^2, \cdots, \sigma_{np}^2, \nonumber\\
& &\ l_2, \cdots, l_q, m_2, \cdots, m_q]^T. \nonumber
\end{eqnarray}
In this parameter vector, $\phi_1$ and $\sigma_1^2$ are omitted because they
are set to constants for the problem to be identifiable.  Indeed, the
restriction $\sigma_1^2 = 1$ is imposed by the fact that $\mathbf{G}$ and
$\bSigma$ share a common factor, while the first constraint is required since
one can only measure the gain phases with respect to some reference, here
achieved by setting $\phi_1 = 0$.\footnote{ In \cite{Wijnholds2006-2} it is
  shown that $\sum_{i=1}^p \phi_i = 0$ is the optimal constraint for this
  problem. This constraint has the disadvantage that the location of the phase
  reference is not well defined. Furthermore, the choice for the constraint
  used here simplifies our analysis in combination with the constraints
  required to uniquely identify the source locations and the apparent source
  powers.  }

Similarly to \eqref{eq:sigmalsq}, the parameters are obtained by solving
\begin{equation}
\widehat{\btheta} = \underset{\btheta}{\mathrm{argmin}} \parallel \mathbf{W}_c
\left ( \widehat{\mathbf{R}} - \mathbf{GA} \bSigma \mathbf{A}^H \mathbf{G}^H -
\bSigma_n \right ) \mathbf{W}_c \parallel_F^2,
\label{eq:selfcalimage}
\end{equation}
where $\mathbf{G}$, $\mathbf{A}$, $\mathbf{\Sigma}$ and $\mathbf{\Sigma}_n$
are all functions of $\btheta$ and $\mathbf{W}_c = \mathbf{R}^{-\frac{1}{2}}
\approx \frac{1}{\sigma_n} \mathbf{I}$ as argued earlier.

The minimum variance for an unbiased estimator is given by the
Cram\`er-Rao Bound (CRB). The CRB on the error variance for any unbiased
estimator states that the covariance matrix $\mathbf{C}_\theta$ 
of the parameter vector $\btheta$ satisfies \cite{Kay1993-1}
\begin{equation}
\mathbf{C}_\theta 
= \expect \left \{ \left ( \widehat{\btheta} - \btheta \right )
  \left ( \widehat{\btheta} - \btheta \right )^T \right \} \geq \frac{1}{N}
\mathbf{J}^{-1},
\end{equation}
where $\mathbf{J}$ is the Fisher information matrix (FIM).
For Gaussian data models $\mathbf{J}$ can be expressed as
(e.g. \cite{Stoica1996-1})
\begin{equation}
\mathbf{J} = \mathbf{F}^H \left ( \overline{\mathbf{R}}^{-1} \otimes
  \mathbf{R}^{-1} \right ) \mathbf{F}
\end{equation}
where $\mathbf{R}$ is the data covariance matrix and $\mathbf{F}$ is 
the Jacobian evaluated at the true values of the
parameters, i.e.,
\begin{equation}
\mathbf{F} = \frac{\delta \vec ( \mathbf{R} )}{\delta \btheta^T} \Big
  |_{\btheta}. \label{eq:Fdefinition}
\end{equation}

For the self calibration scenario, the Jacobian can be partitioned into six
parts following the structure of $\btheta$:
\begin{equation}
\mathbf{F} = [\mathbf{F}_\gamma, \mathbf{F}_\phi, \mathbf{F}_\sigma,
\mathbf{F}_{\sigma_n}, \mathbf{F}_l, \mathbf{F}_m].
\end{equation}
By substitution of \eqref{eq:R2} in \eqref{eq:Fdefinition}, it follows
directly that the first four components can be expressed as
\begin{eqnarray}
\mathbf{F}_\gamma & = & \left ( \overline{\mathbf{G} \mathbf{R}_0} \bPhi
\right ) \circ \mathbf{I} + \mathbf{I} \circ \left ( \mathbf{G} \mathbf{R}_0
  \overline{\bPhi} \right )
\label{eq:Fgamma}\\
\mathbf{F}_\phi & = & \jcmplx \left ( \left ( \overline{\mathbf{G}
      \mathbf{R}_0} \mathbf{G} \right ) \circ \mathbf{I} - \mathbf{I} \circ
  \left ( \mathbf{G} \mathbf{R}_0 \overline{\mathbf{G}} \right ) \right )
\mathbf{I}_s
\label{eq:Fphi}\\
\mathbf{F}_\sigma & = & \left ( \left ( \overline{\mathbf{G} \mathbf{A}}
  \right ) \circ \left ( \mathbf{G} \mathbf{A} \right ) \right )
\mathbf{I}_s
\label{eq:Fsigma}\\
\mathbf{F}_{\sigma_n} &=& \mathbf{I} \circ \mathbf{I}
\label{eq:Fsigman}
\end{eqnarray}
where $\mathbf{R}_0 = \mathbf{A} \bSigma \mathbf{A}^H$ and $\mathbf{I}_s$ is a
selection matrix of appropriate size equal to the identity matrix with its
first column removed so that the derivatives with respect to $\phi_1$ and
$\sigma_1^2$ are omitted.

If the receiver noise powers of all $p$ elements are the same, the
expression for $\mathbf{F}_{\sigma_n}$ given in \eqref{eq:Fsigman} should
be replaced by
\begin{equation}
\mathbf{F}_{\sigma_n}  =  \vec ( \mathbf{I} ) 
\label{eq:Fsigmanscalar}
\end{equation}

For the last two components of the FIM, derivatives of $\vec \left (
  \mathbf{R} \right )$ with respect to the source position coordinates are
required. Let $(x_i,y_i)$ be the coordinates of the $i$th array element,
and introduce 
\begin{eqnarray}
\mathbf{G}_x & = & \diag([x_1, x_2, \cdots x_p]^T) \mathbf{G}\\
\mathbf{G}_y & = & \diag([y_1, y_2, \cdots y_p]^T) \mathbf{G}
\end{eqnarray}
then these components can be conveniently written as
\begin{eqnarray}
\mathbf{F}_l & = & -\jcmplx \frac{2\pi}{\lambda} \left ( \overline{\mathbf{G}
    \mathbf{A}} \circ \mathbf{G}_x \mathbf{A} - \overline{\mathbf{G}_x
    \mathbf{A}} \circ \mathbf{G} \mathbf{A} \right ) \bSigma \mathbf{I}_s\\
\mathbf{F}_m & = & -\jcmplx \frac{2\pi}{\lambda} \left ( \overline{\mathbf{G}
    \mathbf{A}} \circ \mathbf{G}_y \mathbf{A} - \overline{\mathbf{G}_y
    \mathbf{A}} \circ \mathbf{G} \mathbf{A} \right ) \bSigma \mathbf{I}_s
\end{eqnarray}

These equations show that the entries of the Jacobian related to derivatives
with respect to the $l$- and $m$-coordinates of the sources are proportional
to the $x$- and $y$-coordinates of the array elements respectively. The
physical interpretation of this relation is that a plane wave propagating
along the coordinate axis of the coordinate to be estimated provides a more
useful test signal to estimate the source location than a signal propagating
perpendicular to this axis.

The preceding equations allow us to compute $\mathbf{C}_\theta$. The variance
of the estimated image values, i.e. the noise on the image values due to
estimation inaccuracy, is given by the diagonal of the sub-block
$\mathbf{C}_{\sigma\sigma}$ of this matrix, following the partitioning of
$\btheta$.  In general $\mathbf{C}_{\sigma\sigma}$ is not a diagonal
matrix. The other entries in this sub-block describe the way in which the
noise on the pixels are correlated among themselves---this is associated with
false structures.

\subsection{Propagation of calibration errors}
\label{sec:errprop}

If the instrumental parameters are extracted from separate calibration data,
the minimum variance on these estimated values is given by the CRB on the
instrumental parameters in the calibration experiment, $\mathbf{C}_\theta$,
where now $\btheta = [\gamma_1, \cdots, \gamma_p, \phi_2, \cdots, \phi_p,
\sigma_{n1}^2, \cdots, \sigma_{np}^2]^T$.  With this choice for $\btheta$ the
results for $\mathbf{F}_\gamma$, $\mathbf{F}_\phi$ and $\mathbf{F}_{\sigma_n}$
derived earlier in (\ref{eq:Fgamma}), (\ref{eq:Fphi}) and (\ref{eq:Fsigman})
can be used assuming that the calibration measurement adheres to the same data
model. The propagation of the calibration errors to the image is described by
\begin{equation}
\mathrm{cov} \left ( \mathbf{i} \right ) = \left ( \frac{\partial
    \mathbf{i}}{\partial \btheta^T} \right ) \mathbf{C}_\theta 
    \left ( \frac{\partial \mathbf{i}}{\partial \btheta^T} \right)^T.
\label{eq:errprop}
\end{equation}
We thus need to derive $\partial \mathbf{i}/\partial \bgamma^T$, $\partial
\mathbf{i}/\partial \bphi^T$ and $\partial \mathbf{i}/\partial \bsigma_n^T$.

The derivative of the image values to $\gamma_k$ is defined as
\begin{equation}
  \frac{\partial \mathbf{i}}{\partial \gamma_k} = \frac{\partial}{\partial
    \gamma_k} \mathbf{M}^{-1} \left ( \overline{\mathbf{GA}} \circ \mathbf{GA}
  \right )^H \vec \left ( \mathbf{R} - \bSigma_n \right )
\end{equation}
where $\mathbf{M}$ and $\mathbf{G}$ depend on $\mathbf{\gamma}$.
Applying the formula for the derivative of an inverted matrix with respect to
one of its elements \cite{Moon2000-1}, this can be rewritten as
\begin{eqnarray}
\lefteqn{\frac{\partial \mathbf{i}}{\partial \gamma_k} =} \nonumber\\
& &= \Bigg ( -\mathbf{M}^{-1} \left ( \frac{\partial}{\partial \gamma_k}
  \mathbf{M}  \right ) \mathbf{M}^{-1} \left ( \overline{\mathbf{GA}} \circ
  \mathbf{GA} \right )^H + \nonumber\\
& &+ \mathbf{M}^{-1} \gamma_k \left ( \e^{-\jcmplx \phi_k} \mathbf{E}_{kk}
  \overline{\mathbf{A}} \circ \mathbf{GA} + \overline{\mathbf{GA}} \circ
  \e^{\jcmplx \phi_k} \mathbf{E}_{kk} \mathbf{A} \right )^H \Bigg )
\nonumber\\
& &\times \vec \left ( \mathbf{R} - \bSigma_n \right ) \label{eq:didgamma1}
\end{eqnarray}
where $\mathbf{E}_{kk}$ is the elementary matrix with all its entries set to
zero except element $E_{kk}$ which is set to 1. Inserting the vectorized
version of \eqref{eq:R2} in \eqref{eq:didgamma1} and removing the
Khatri-Rao products, we obtain
\begin{equation}
\frac{\partial \mathbf{i}}{\partial \gamma_k} = -2 \gamma_k \left ( 2 -
  \gamma_k \right ) \mathbf{M}^{-1} \mathrm{Re} \left \{
  \overline{\mathbf{A}}_{k:}^H \overline{\mathbf{A}}_{k:} \odot \mathbf{A}^H
  \bGamma^2 \mathbf{A} \right \} \bsigma \label{eq:didgamma2}
\end{equation}
We have introduced the notation $\mathbf{A}_{k:} = \mathbf{E}_{kk}
\mathbf{A}$, i.e. $\mathbf{A}_{k:}$ has only zero valued entries except on the
$k$th row where the elements are equal to the corresponding elements
of $\mathbf{A}$. The goal of this derivation is to obtain an expression for
$\partial \mathbf{i} / \partial \bgamma^T$. We will thus have to stack the
expression for $\partial \mathbf{i} / \partial \gamma_k$ in a single
matrix. This is facilitated by introducing $\mathbf{a}_k$ as the
$k$th row of $\mathbf{A}$ and rewriting \eqref{eq:didgamma2} as
\begin{equation}
\frac{\partial \mathbf{i}}{\partial \gamma_k} = -2\gamma_k \left (2 -
  \gamma_k \right ) \mathbf{M}^{-1} \mathrm{Re} \left \{ \mathbf{a}_k^T
  \mathbf{1}^T \odot \mathbf{A}^H \bGamma^2 \mathbf{A} \bSigma \mathbf{a}_k^H
\right \}
\end{equation}
where $\mathbf{1}$ denotes a vector of ones of appropriate size.

By stacking all vector $\partial \mathbf{i} / \partial \gamma_k$ in a single
matrix, we thus obtain
\begin{equation}
\frac{\partial \mathbf{i}}{\partial \bgamma^T} = -2 \mathbf{M}^{-1}
\mathrm{Re} \left \{ \mathbf{A}^T \odot \mathbf{A}^H \bGamma^2 \mathbf{A}
  \bSigma  \mathbf{A}^H \right \} \left ( 2 \mathbf{I} - \bGamma \right )
\bGamma. \label{eq:didgamma3}
\end{equation}

The corresponding result for $\phi_k$ can be derived in a similar way, so we
only present the main steps.
\begin{eqnarray}
\lefteqn{\frac{\partial \mathbf{i}}{\partial \phi_k} =} \nonumber\\
& = & \frac{\partial}{\partial \phi_k} \left ( \overline{\mathbf{G} \mathbf{A}
  } \circ \mathbf{G} \mathbf{A} \right )^\dagger \vec \left ( \mathbf{R} \right ) \nonumber\\
& = & \mathbf{M}^{-1} \left ( \frac{\partial}{\partial \phi_k}
  \overline{\mathbf{GA}} \circ \mathbf{GA} \right )^H \left ( \left (
    \overline{\mathbf{G} \mathbf{A}} \circ \mathbf{G} \mathbf{A} \right )^H
  \bsigma \right ) \nonumber\\
& = & \mathbf{M}^{-1} \left ( -\jcmplx \e^{-\jcmplx\phi_k} \bGamma
  \overline{\mathbf{A}}_{k:} \circ \mathbf{GA} + \jcmplx \e^{\jcmplx \phi_k}
  \overline{\mathbf{GA}} \circ \bGamma \mathbf{A}_{k:} \right )^H \nonumber\\
& & \times \left ( \left ( \overline{\mathbf{G} \mathbf{A}} \circ \mathbf{G}
    \mathbf{A} \right )^H \bsigma \right ).
\end{eqnarray}
Removal of the Khatri-Rao products by reducing them to Hadamard products
gives
\begin{eqnarray}
\frac{\partial \mathbf{i}}{\partial \phi_k} & = & -2 \gamma_k^2
\mathbf{M}^{-1} \mathrm{Im} \left \{ \left ( \overline{\mathbf{A}}_{k:}^H
    \overline{\mathbf{A}_{k:}} \odot \mathbf{A}^H \bGamma^2 \mathbf{A} \right ) \right \}
\bsigma.
\end{eqnarray}

Note that this term has the same form as the first term in
\eqref{eq:didgamma2}, so it can be rewritten in a similar way. This gives
\begin{equation}
\frac{\partial \mathbf{i}}{\partial \phi_k} = -2 \gamma_k^2 \mathbf{M}^{-1}
\mathrm{Im} \left \{ \mathbf{a}_k^T \mathbf{1}^T \odot \mathbf{A}^H \bGamma^2
  \mathbf{A} \bSigma \mathbf{a}_k^H \right \}
\end{equation}
and therefore
\begin{equation}
\frac{\partial \mathbf{i}}{\partial \bphi^T} = -2 \mathbf{M}^{-1}
\mathrm{Im} \left \{ \overline{\mathbf{A}}^H \odot \mathbf{A}^H \bGamma^2
  \mathbf{A} \bSigma \mathbf{A}^H \right \} \bGamma^2 \label{eq:didphi}
\end{equation}

Finally, the partial derivative of the image values with respect to
$(\sigma_{nk}^2)$ is given by
\begin{eqnarray}
\frac{\partial \mathbf{i}}{\partial (\sigma_{nk}^2)} & = &
\frac{\partial}{\partial (\sigma_{nk}^2)} \mathbf{M}^{-1} \left (
  \overline{\mathbf{GA}} \circ \mathbf{GA} \right )^H \vec \left ( \mathbf{R}
  - \bSigma_n \right ) \nonumber\\
& = & \mathbf{M}^{-1} \left ( \overline{\mathbf{GA}} \circ \mathbf{GA} \right
)^H \vec \left ( -\mathbf{E}_{kk} \right ) \nonumber\\
& = & -\mathbf{M}^{-1} \left ( \left | \mathbf{GA} \right |^{\odot 2} \right
)^H \vecdiag \left ( \mathbf{E}_{kk} \right ).
\end{eqnarray}
Therefore
\begin{equation}
\frac{\partial \mathbf{i}}{\partial \bsigma_n^T} = -\mathbf{M}^{-1} \left (
  \left | \mathbf{GA} \right |^{\odot 2} \right )^H. \label{eq:didsigmanvec}
\end{equation}
If $\bSigma_n = \sigma_n^2 \mathbf{I}$ this reduces further to
\begin{equation}
\frac{\partial \mathbf{i}}{\partial \sigma_n} = -\mathbf{M}^{-1} \left ( \left
    | \mathbf{GA} \right |^{\odot 2} \right )^H
\mathbf{1}. \label{eq:didsigmanscalar}
\end{equation}

The partial derivatives as well as the CRB \cite{Wijnholds2006-3} contain
terms involving $\mathbf{A}^H \mathbf{A}$, often weighted by the gains of the
receiving elements. Given the physical interpretation of this factor discussed
in section \ref{sec:imaging}, this suggests that the error patterns introduced
in the image by calibration errors follow the structures in the dirty
image. This is confirmed by the example in section
\ref{sec:implications}. Since the CRB is inversely proportional with $N$,
which is equal to the product of bandwidth and integration time, the image
covariance due to calibration errors decreases proportional to bandwidth and
integration time.

\subsection{Thermal noise}

In this section we derive an expression for the covariance of the image values
due to the thermal noise in the data. We will therefore assume that perfect
knowledge of the thermal noise power $\bSigma_n$ is available to avoid
confusion between the thermal noise contribution and the contribution of
propagated estimation errors. The covariance of the image values is by
definition given by
\begin{eqnarray}
\lefteqn{\mathrm{cov} \left ( \mathbf{i} \right ) =} \nonumber\\
& = & \expect \left \{ \left ( \vec \left ( \widehat{\mathbf{i}} \right ) -
    \vec \left ( \mathbf{i} \right ) \right ) \left ( \vec \left (
      \widehat{\mathbf{i}} \right ) - \vec \left ( \mathbf{i} \right ) \right
  )^H \right \} \nonumber\\
& = & \expect \Big \{ \left ( \overline{\mathbf{GA}} \circ \mathbf{GA} \right
)^\dagger \left ( \vec \left ( \widehat{\mathbf{R}} - \bSigma_n \right ) -
  \vec \left ( \mathbf{R} - \bSigma_n \right ) \right ) \nonumber\\
& & \times \left ( \vec \left ( \widehat{\mathbf{R}} - \bSigma_n \right ) -
  \vec \left ( \mathbf{R} - \bSigma_n \right ) \right )^H \nonumber\\
& & \times \left ( \overline{\mathbf{GA}} \circ \mathbf{GA} \right )^{\dagger
  H} \Big \}.
\end{eqnarray}
This shows that under the assumption that perfect knowledge on
$\bSigma_n$ in $\widehat{\mathbf{R}}$ is available, $\bSigma_n$ drops out. Furthermore,
$(\overline{\mathbf{GA}} \circ \mathbf{GA})^\dagger$ can be moved outside the
expectation operator, since it contains no estimated values. Therefore
\begin{equation}
\mathrm{cov} \left ( \mathbf{i} \right ) = \mathbf{M}^{-1} \left (
  \overline{\mathbf{GA}} \circ \mathbf{GA} \right )^H \mathrm{cov} \left (
  \mathbf{R} \right ) \left ( \overline{\mathbf{GA}} \circ \mathbf{GA} \right
) \mathbf{M}^{-1}.
\end{equation}
For Gaussian data models
\begin{equation}
\mathrm{cov} \left ( \mathbf{R} \right ) = \frac{1}{N} \left (
  \overline{\mathbf{R}} \otimes \mathbf{R} \right ) ,
\end{equation}
and we find that
\begin{eqnarray}
\lefteqn{\mathrm{cov} \left ( \mathbf{i} \right ) =} \nonumber\\
& = & \frac{1}{N} \mathbf{M}^{-1} \left ( \overline{\mathbf{GA}} \circ
  \mathbf{GA} \right )^H \left ( \overline{\mathbf{R}} \otimes \mathbf{R}
\right ) \left ( \overline{\mathbf{GA}} \circ \mathbf{GA} \right )
\mathbf{M}^{-1}.\nonumber
\end{eqnarray}
This can be rewritten using Kronecker and Khatri-Rao product relations as
\begin{eqnarray}
\mathrm{cov} \left ( \mathbf{i} \right )  & = & \frac{1}{N} \mathbf{M}^{-1}
\left ( \overline{\mathbf{GA}} \circ \mathbf{GA} \right )^H \left (
  \overline{\mathbf{RGA}} \circ \mathbf{RGA} \right) \mathbf{M}^{-1}
\nonumber\\
  & = & \frac{1}{N} \mathbf{M}^{-1} \left | \mathbf{A}^H \mathbf{G}^H
    \mathbf{RGA} \right |^{\odot 2} \mathbf{M}^{-1}.
\end{eqnarray}
Finally, substituting the data model presented in \eqref{eq:R2} we get
\begin{eqnarray}
\lefteqn{ \mathrm{cov} \left ( \mathbf{i} \right ) =} \nonumber\\
& = & \frac{1}{N} \mathbf{M}^{-1} \left | \mathbf{A}^H \bGamma^2 \mathbf{A}
  \bSigma \mathbf{A}^H \bGamma^2 \mathbf{A} + \mathbf{A}^H \bGamma^2 \bSigma_n
  \mathbf{A} \right |^{\odot 2} \mathbf{M}^{-1} \nonumber\\ \label{eq:covnoise}
\end{eqnarray}

It is interesting to note that for an array having $\mathbf{G} = \mathbf{I}$ a
diagonalization of $\mathbf{A}^H \mathbf{A}$ does not only ensure a
homogeneous noise distribution over the map after the deconvolution operation
as demonstrated in Sec. \ref{ssec:noiseredistribution}, but also diagonalizes
the image covariance due to thermal noise, thus ensuring that the noise on the
pixels is uncorrelated. The Gram matrix $\mathbf{A}^H \mathbf{A}$ describes
the amount of linear independence (or orthogonality) of the direction of
arrival vectors within the field of view of the array, which can be visualized
as the array beam pattern. This observation therefore suggests that an array
with a low side lobe pattern does not only provide good spatial separation
between source signals, but also gives small covariance between image values
after deconvolution.

\subsection{Confusion noise}

The contributions to the effective noise from calibration errors and thermal
noise scale inversely with the number of samples $N$, which is equal to the
product of bandwidth and integration time. This implies that, theoretically,
these sources of image noise can be reduced to arbitrarily low levels. In
practice, the radio astronomical array will detect more sources with every
reduction of the noise in the map. At some point, the number of detected
sources becomes larger than the number of resolution elements in the image,
which will turn the map into one blur of sources. The maximum density of
discernable sources is the {\em classical confusion limit} and relates to the
resolution of the image.

In terms of self calibration, to have more detectable sources requires more
source parameters to describe the source model. At some point, the self
calibration problem becomes ill-posed. We will refer to this as the self
calibration confusion limit. Although the exact limit depends on the minutiae
of the array and source configuration, we can easily compute an upper limit on
the tractable number of sources based on the argument that the number of
unknowns should be smaller than the number of equations. The data model
provides a relation between the parameters and the data. For a $p$ element
array, the covariance matrix contains $p^2$ independent real values, so the
data model can be regarded as $p^2$ independent equations. Solving for the
direction independent complex gains requires $2p-1$ real valued parameters,
estimation of the receiver element noise powers requires another $p$
parameters and the $q$ sources are described by $3q-1$ parameters (the
apparent source powers relative to the first source and two coordinates per
source). The self calibration problem is therefore constrained by
\begin{equation}
p^2 \geq 2p-1 + p + 3q - 1 = 3p + 3q - 2
\end{equation}
implying that
\begin{equation}
q \leq \frac{p^2 - 3p + 2}{3}.
\label{eq:limit}
\end{equation}

The spatial Nyquist sampling with the $8\times 8$ URA allows an image grid of
$15 \times 15 = 225$ image values. This resolution was confirmed by the
condition number analysis presented in
Fig. \ref{fig:conditionnumber}. However, the upper limit based on the analysis
above for a 64 element array is 1302. The mismatch between this upper limit
and the actual number of uniquely solvable image values can be attributed to
the redundancy in the array. Due to this redundancy the cross-correlations of
many antenna pairs provide the same spatial information instead of providing
additional information on the spatial structure of the sky. In terms of the
argument leading to Eq. \eqref{eq:limit}, there is linear dependence between
the equations and therefore the number of equations that can be used to solve
parameters is reduced. The 5-armed array performs much better in this
regard. E.g., for $p = 40$ the upper limit on the number of sources given by
(\ref{eq:limit}) is 494. Since $\sqrt{494} \approx 22$, we can thus form an
image grid of $22 \times 22$ points, thus providing a resolution of $\Delta l
\approx 0.09$. Figure \ref{fig:conditionnumber} shows that the condition
number for this array goes to numerical infinity at $\Delta l \approx 0.08$,
showing that the 5-armed array approaches its theoretical self calibration
confusion limit. This example illustrates that if processing power is cheap
compared to antenna hardware, a non-redundant array should be preferred over a
redundant array if the confusion limit should be reached without introducing
an ill-posed deconvolution problem.

In this section we addressed the classical confusion limit for pure imaging
problems and the self calibration confusion limit in self calibration
problems. This type of confusion is often called {\em source confusion} as
opposed to {\em side lobe confusion} which refers to blurring of the image by
side lobe leftovers introduced in the CLEAN process. In the analysis of this
paper, side lobe confusion is part of the deconvolution problem and is thus
intrinsically included in the analysis of calibration error and thermal noise
propagation, and does not need to be addressed separately.  Source confusion
does require a separate treatment because it involves the source density
distribution as function of source brightness.

\section{Implications}
\label{sec:implications}

\subsection{Thermal noise vs.\ propagated calibration errors}

\begin{figure}
\centering
\includegraphics[width=8.5cm]{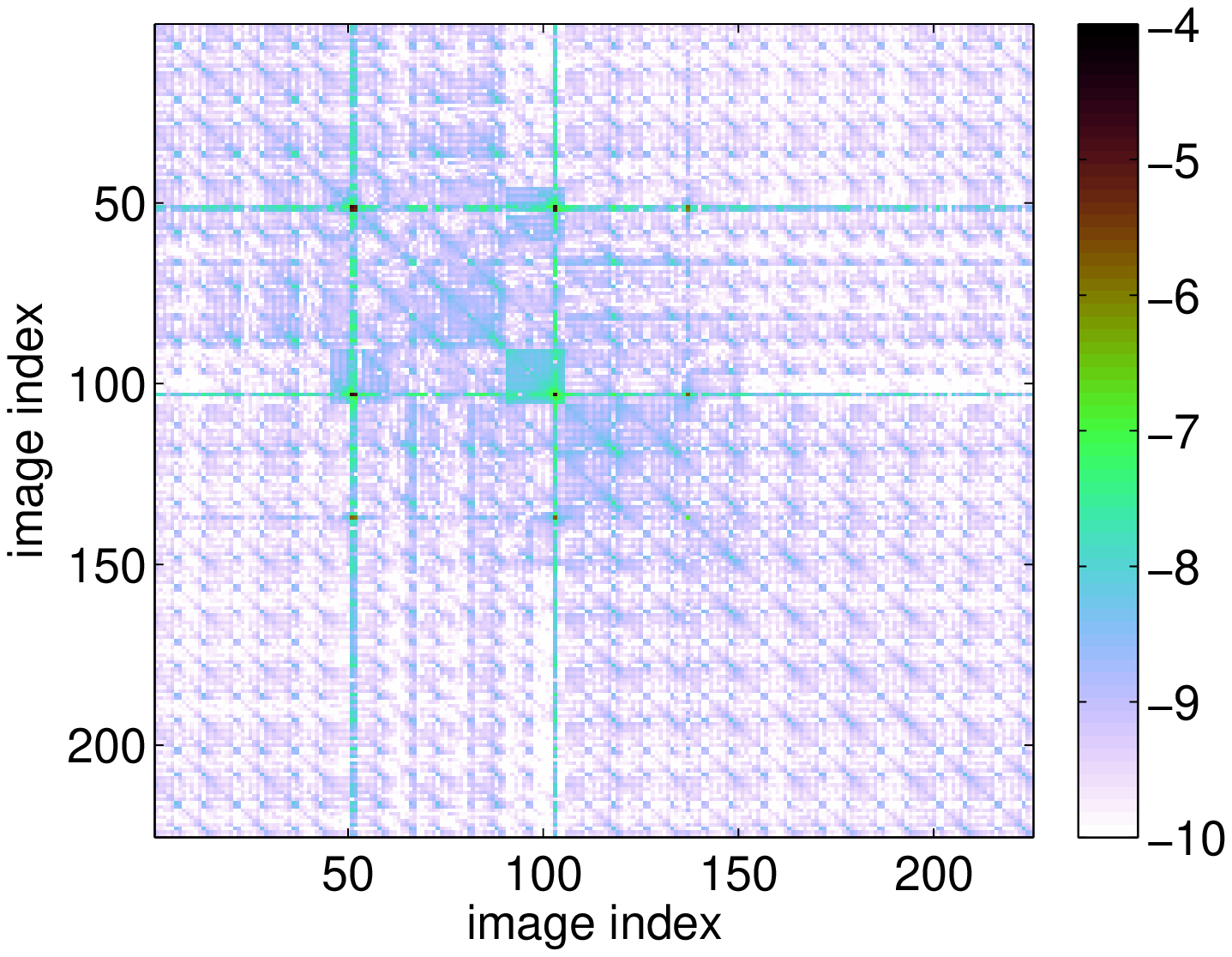}
\includegraphics[width=8.5cm]{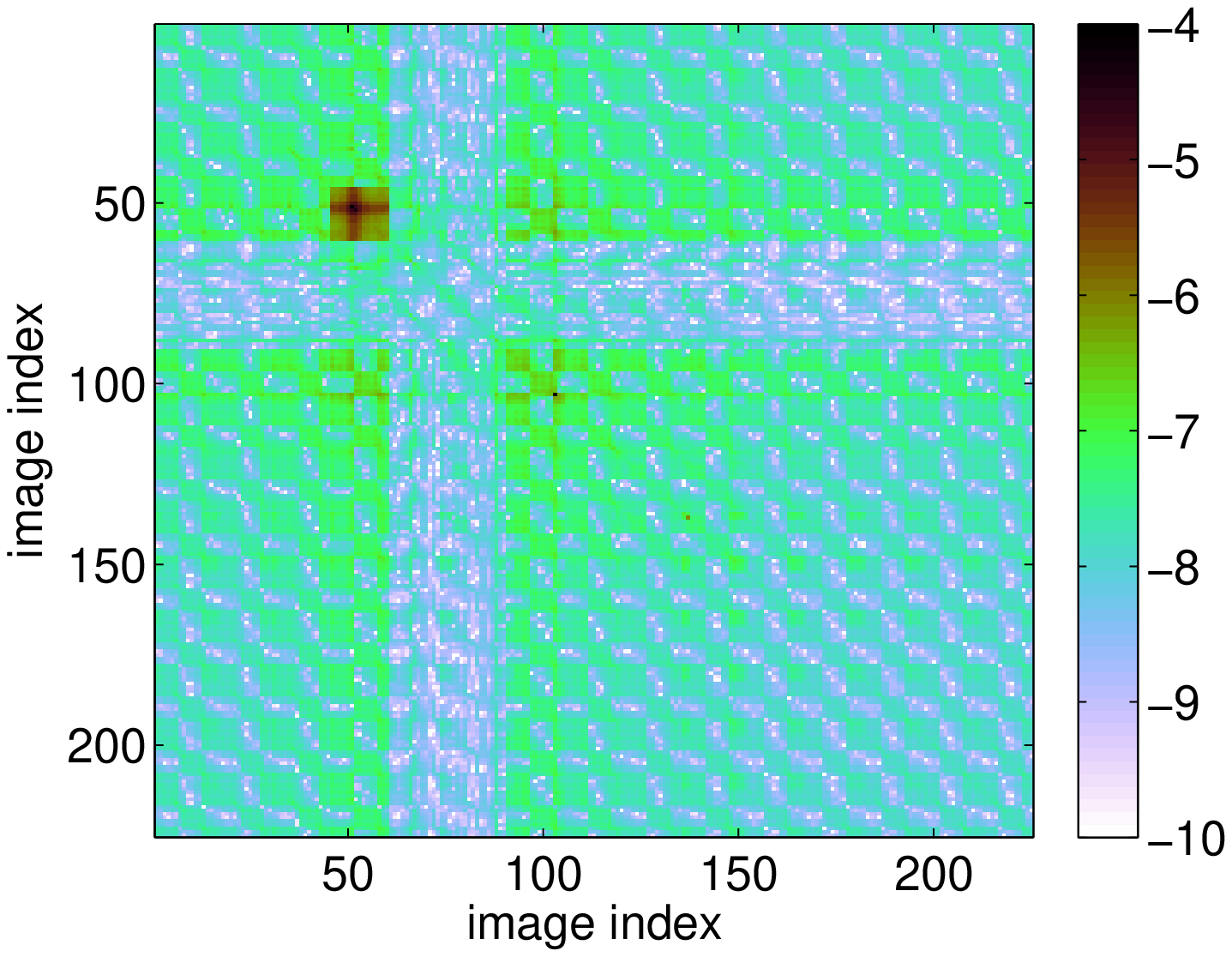}
\caption{($a$) The logarithm (base 10) of the image covariance matrix due
  to calibration errors and ($b$) due to the measurement noise on the same
  color scale. The former is almost everywhere two orders of magnitude lower.
  \label{fig:imcov}}
\end{figure}

We compare the image covariance due to calibration errors to the image
covariance due to the noise on the data in a simulation. The calibration
parameters are calculated from a separate data set with the same data model
and integration time. We computed the CRB for the $8\times 8$ URA using the
relations presented in Section \ref{sec:CRB} for the simultaneous solution of
the omni-directional complex gains, $\mathbf{G}$, and the system noise power,
$\sigma_n^2$, which was assumed to be the same for all array elements,
assuming a short term integration over $N = 16384$ samples. This CRB was used
in \eqref{eq:errprop} to compute the image covariance matrix due to
calibration errors. The magnitudes of the matrix entries are shown in Fig.\
\ref{fig:imcov}($a$), in a log scale.

The image covariance matrix due to the noise in the measurement was calculated
using \eqref{eq:covnoise} and is shown in Fig.\ \ref{fig:imcov}($b$). This
shows that the covariance of the image values due to the calibration errors is
more concentrated at the source locations than the covariance due to the
system noise, but is generally more than two orders of magnitude lower. These
results indicate that the calibration errors only represent a minor
contribution to the total effective noise, even when the calibration
measurements are done on the same (short) time scales using sources of the
same strength, i.e. when the calibration measurement is similar to the actual
measurement.

\subsection{Calibration observations vs.\ self calibration}

In the previous section we discussed the situation in which the array is
calibrated in a separate measurement. This scheme requires an extremely stable
instrument. In most practical applications, the calibration is therefore done
on the same data that is also used to provide the final image (self
calibration). It is interesting to see how these scenarios compare. To this
end, we used the relationships presented in Section \ref{sec:CRB} to compute
the CRB for simultaneous estimation of the omni-directional complex gains, the
apparent source powers, the source locations and the system noise power for
the $8\times 8$ URA.

\begin{figure}
\centering
\includegraphics[width=8.5cm]{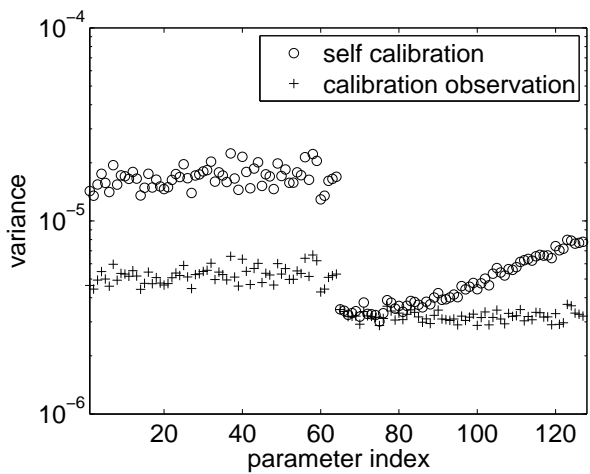}
\caption{The CRB for the omni-directional complex gain
  amplitudes (parameters 1 through 64) and phases (parameters 65 through 127)
  for a separate calibration observation and the self calibration
  approach. \label{fig:CRBgain}}
\end{figure}

Figure \ref{fig:CRBgain} compares the CRBs for these two cases.  The expected
covariance of the gains and phases in the self calibration experiment is
higher since more parameters have to be estimated simultaneously. The behavior
of the CRB on the phases in the self calibrated observation (sloped upwards
for increasing parameter index) can be explained by the interaction between
the source parameters and the gain phases combined with the choice of the
phase reference element in the corner of the array.

\begin{table}
\centering
\caption{Covariance of source power estimates.\label{tab:covsrc}}
\begin{tabular}{lrrr}
\multicolumn{4}{c}{self calibration}\\
\hline
index & 2 & 3 & 4\\
2 & $0.266 \times 10^{-4}$ & $0.287 \times 10^{-4}$ & $0.022 \times 10^{-4}$\\
3 & $0.287 \times 10^{-4}$ & $1.237 \times 10^{-4}$ & $0.048 \times 10^{-4}$\\
4 & $0.022 \times 10^{-4}$ & $0.048 \times 10^{-4}$ & $0.007 \times 10^{-4}$\\
\hline
\hline\\[3ex]
\multicolumn{4}{c}{separate calibration}\\
\hline
index & 2 & 3 & 4\\
2 & $0.280 \times 10^{-4}$ & $0.072 \times 10^{-4}$ & $0.006 \times 10^{-4}$\\
3 & $0.072 \times 10^{-4}$ & $0.772 \times 10^{-4}$ & $0.012 \times 10^{-4}$\\
4 & $0.006 \times 10^{-4}$ & $0.012 \times 10^{-4}$ & $0.005 \times 10^{-4}$\\
\hline
\hline
\end{tabular}
\end{table}

Table \ref{tab:covsrc} shows, for each of the two cases, the covariance
matrices of the apparent source powers, i.e., the variance of the image values
at the locations of the sources.  The scaling factor ambiguity between
$\mathbf{G}$ and $\bSigma$ in the self calibration case is resolved by putting
$\sigma_1^2 = 1$ to constrain the problem, and therefore only the covariance
values of the other three sources is tabulated.  For the case with a separate
calibration observation the covariance matrix was extracted from the sum of
the image covariances due to calibration errors and system noise.  The results
in the table indicate that the variance of the source power estimates in both
cases are comparable, although the source power estimates are slightly better
when gain calibration data is available from a separate measurement. The
covariance values found for a separate calibration stage are much lower than
the corresponding values for self calibration. This suggests that pure imaging
is more capable of separating source signals from different directions than
self calibrated imaging.

\section{Conclusions}

In this paper we presented an analytic solution for snapshot imaging including
deconvolution based on a data model (measurement equation) for the antenna
signal covariance matrix or visibilities. The presented comprehensive
framework is sufficiently flexible to enable extension of this analysis to
synthesis observations, since the data model for a synthesis observation has
the same form \cite{Leshem2000-1, Veen2004-1, Tol2007-1}. This framework
allowed us to make the first complete rigorous assessment of the effective
noise floor, which is the combined effect of propagated calibration errors,
thermal noise and source confusion, in the image in terms of the covariance of
the image values. Our simulations for a 2D uniform rectangular array indicate
that the effect of propagated calibration errors is strongly concentrated at
the source locations but is considerably smaller than the thermal noise at
other image points. The results also suggest that if the instrument is
sufficiently stable, a separate calibration step is to be preferred over a
self calibrated image since it allows better source separation in the imaging
process.

The effects of deconvolution can be described by a deconvolution matrix that
describes the amount of linear independence (orthogonality) of the spatial
signature vectors weighted by the actual gains of the receiving elements. A
diagonal deconvolution matrix not only ensures the best possible spatial
separation between the sources, but also ensures a homogeneous noise
distribution over the map. This poses the question whether this matrix can be
diagonalized by array design or by applying appropriate weights to the array
elements. Since this factor is related to the array beam pattern, the latter
is equivalent to finding weights that suppress the side lobe patterns at least
in the direction of other sources, which suggest that techniques like Robust
Capon Beamforming should provide the requested weighting \cite{Tol2005-1}. The
condition number of the deconvolution matrix can be used to assess the quality
of the solution to the deconvolution problem.

Compared to a redundant array (ULA, URA), an array without redundant element
spacings provides much better possibilities to approach the maximum number of
solvable image points for a fixed number of antenna elements, thereby allowing
the system to reach the theoretical self calibration confusion limit.

\bibliography{refs}

\begin{biography}{Stefan J. Wijnholds}
  (S'2006) was born in The Netherlands in 1978. He received M.Sc.\ degrees in
  Astronomy and Applied Physics (both cum laude) from the University of
  Groningen in 2003. After his graduation he joined R\&D department of ASTRON,
  the Netherlands Foundation for Research in Astronomy, in Dwingeloo, The
  Netherlands, where he works with the system design and integration group on
  the development of the next generation of radio telescopes. Since 2006 he is
  also with the Delft University of Technology, Delft, The Netherlands, where
  he is pursuing a Ph.D.\ degree. His research interests lie in the area of
  array signal processing, specifically calibration and imaging.
\end{biography}

\begin{biography}{Alle-Jan van der Veen}
  (F'2005) was born in The Netherlands in 1966. He received the Ph.D.\ degree
  (cum laude) from TU Delft in 1993. Throughout 1994, he was a postdoctoral
  scholar at Stanford University.  At present, he is a Full Professor in
  Signal Processing at TU Delft.

  He is the recipient of a 1994 and a 1997 IEEE Signal Processing Society
  (SPS) Young Author paper award, and was an Associate Editor for IEEE Tr.\
  Signal Processing (1998--2001), chairman of IEEE SPS Signal Processing for
  Communications Technical Committee (2002-2004), and Editor-in-Chief of IEEE
  Signal Processing Letters (2002-2005).  He currently is Editor-in-Chief of
  IEEE Transactions on Signal Processing, and member-at-large of the Board of
  Governors of IEEE SPS.

  His research interests are in the general area of system theory applied to
  signal processing, and in particular algebraic methods for array signal
  processing, with applications to wireless communications and radio
  astronomy.
\end{biography}

\end{document}